\documentclass[reprint,amsmath,amssymb,aps]{revtex4-2}
\usepackage{graphicx}
\usepackage{dcolumn}
\usepackage{bm}
\usepackage{hyperref}
\usepackage[utf8]{inputenc}
\usepackage{amsmath,amssymb}
\usepackage{amsthm}
\usepackage{hyperref}
\usepackage{acronym}
\usepackage{color}
\usepackage{amsfonts}
\usepackage{mathrsfs}
\usepackage{graphicx}
\usepackage[on]{psfrag}
\usepackage{array}
\usepackage{cases}
\usepackage{tabularx}
\usepackage{url}
\usepackage{soul}
\usepackage{theoremref}
\usepackage{thmtools}
\usepackage{lettrine}
\usepackage{fancyhdr}
\usepackage{physics}
\begin{document}

\preprint{APS/123-QED}

\title{Ionization energies in lithium and boron atoms using the Variational Quantum Eigensolver algorithm}

\author{$^1$Rene Villela, $^2$V. S. Prasannaa, and $^{1,2}$ B. P. Das }
\affiliation{$^1$Department of Physics, Tokyo Institute of Technology, 2-12-1-H86 Ookayama, Meguro-ku, Tokyo 152-8550, Japan \\
$^2$Centre for Quantum Engineering, Research and Education, TCG CREST, Salt Lake, Kolkata 700091, India }

\date{\today}

\begin{abstract}
  The classical-quantum hybrid Variational Quantum Eigensolver algorithm is the most widely used approach in the Noisy Intermediate Scale Quantum era to obtain ground state energies of atomic and molecular systems. In this work, we extend the scope of properties that can be calculated using the algorithm by computing the first ionization energies of Lithium and Boron atoms. We check the precision of our ionization energies and the observed many-body trends and compare them with the results from calculations carried out on traditional computers. 
\end{abstract}

\maketitle

\section{Introduction}

\quad Recent advances in quantum sciences and technologies have placed us in the midst of the second quantum revolution~\cite{sqr}. In particular, there is growing interest in the quantum many body theory of atoms and molecules using quantum computers, in view of the quantum devices promising an exponential speed-up in the otherwise very expensive atomic and molecular calculations on traditional computers~\cite{feynman,AandL1,AandL2,TS}. Quantum algorithms such as the phase estimation approach can become prohibitively expensive in the current Noisy Intermediate Scale Quantum (NISQ) era, where we are limited by the number of qubits and noise~\cite{AandL2,QPE,lanyon, Mh, bay}, whereas the classical-quantum hybrid variational quantum eigensolver (VQE) algorithm is well-suited for it~\cite{mhyung,peruzzo}. \\

So far, the VQE algorithm has been employed mainly to calculate ground state energies of closed-shell atomic systems and molecules, which is perhaps its most direct and straightforward application. Therefore, one finds a fairly large number of papers on ground state energies (for example, see Refs.~\cite{qconqc1,qconqc2,qconqc3,qconqc4,qconqc5,qconqc6}), but not nearly as many on other properties, for example, Ref.~\cite{pol}. In this work, we extend the scope of VQE to finding the first ionization energies (IEs) of Lithium ($Li$) and Boron ($B$). To the best of our knowledge, there is only one work on IEs, which uses the phase estimation algorithm~\cite{sugi}, while we have used the more NISQ-friendly VQE algorithm here. From an application point of view, IEs of atomic systems like $Li$ and $B$ are useful in many fields, with examples including astrophysics~\cite{Bastro}, magnetic fusion research~\cite{pppl}, and plasma sources for electron beams~\cite{pl}. From a many-body theoretic perspective, calculating the IE of a closed-shell neutral atom requires computing two ground state energies, one of the atom in a closed-shell configuration (neutral) and another of a single valence electron (singly ionized). In that sense, calculating this property using a quantum many-body theory serves as a quantitative test of that approach. In fact, a wide range of theories ranging from the traditional coupled cluster method to the equation of motion coupled cluster theory have been employed to calculate this property, for example, see Refs.~\cite{Himadri,bksie} and references therein. Our aim in this study is to report all-electron calculations of IEs of $Li$ and $B$ within the VQE framework using the Unitary Coupled Cluster (UCC) ansatz across four appropriately chosen single-particle basis sets. At this juncture, we add that the coupled cluster theory is regarded as the gold standard of electronic structure calculations~\cite{Bartlett}, and is one of the most powerful methods available for calculating properties of quantum many-electron systems. It is therefore only appropriate to employ the unitary variant of the approach for our VQE computations. As both $Li$ and $B$ are small systems, we can carry out UCC calculations with all the electrons being treated as active and no virtual orbitals being cut-off. This allows one to capture electron correlation effects with the VQE algorithm, to the most extent that the underlying ansatz allows. We can then compare the precision of the obtained results with those from full configuration interaction, which is the best that one can obtain within that single-particle basis. \\ 

The manuscript is organised as follows: Section~\ref{theory} presents the theoretical framework adopted in this work as well as the details of our calculations. This is followed by a discussion of our results in Section~\ref{results}, and we finally conclude in Section~\ref{conclusion}. \\

\section{Theory and methodology}\label{theory}

The Rayleigh-Ritz variational principle ensures that the ground state energy of a many-body system, such as an atom or molecule, that is obtained by minimizing the functional 

\begin{eqnarray}
E(\theta)=\frac{\mel{\Psi(\theta)}{{H}}{\Psi(\theta)}}{\braket*{\Psi(\theta)}{\Psi(\theta)}}, \label{eqE}
\end{eqnarray} 

with respect to a set of parameters, $\{\theta\}$, denoted compactly in the above equation as $\theta$, is an upper bound to the true ground state energy of that system~\cite{Griffiths}. $H$ and $|\Psi(\theta) \rangle$ refer to the Hamiltonian and the parametrized wave function of the many-body system, respectively. This idea lies at the core of the VQE algorithm. \\

The many-body wave function is written as $|\Psi (\theta) \rangle = U(\theta) | \Phi_0 \rangle$, where $U(\theta)$ is a unitary operator, also referred to as the variational form/ansatz. $| \Phi_0 \rangle$ is the reference state on which $U(\theta)$ acts. Eq. (\ref{eqE}) is now 

\begin{eqnarray}
E(\theta)={\bra{\Phi_0} U^\dagger(\theta)}{{H}}{U(\theta)}\ket{\Phi_0}. \label{Eq2}
\end{eqnarray} 

  \begin{table*}[t]
  \caption{\label{tab:table1} Ground state energies of Lithium and Boron atoms and their singly ionized cations, and their respective first ionization energies, using the STO-3G, STO-6G, 3-21G, and the 6-31G basis sets. All the values have been rounded off to the sixth decimal place. All units are in Hartrees. }
  \begin{ruledtabular}
  \begin{tabular}{ccccccc}
   \multicolumn{7}{c}{\textbf{STO-3G}}\\ \hline
   \textbf{Method}&\textbf{Li$^+$}&\textbf{Li}&\textbf{IE}&\textbf{B$^+$}&\textbf{B}&\textbf{IE} \\ \hline
   HF & -7.135448& -7.315526&0.180078& -23.948470& -24.148989& 0.200519\\
   FCI&-7.135654 & -7.315837&0.180183& -24.009815 & -24.189265&0.179450 \\
   UCCSD &-7.135654& -7.315832&0.180178& -24.009626& -24.177841&0.168215\\
   \hline \hline 
   \multicolumn{7}{c}{\textbf{STO-6G}}\\ \hline
   HF & -7.221334& -7.399931&0.178597& -24.190562&-24.394295& 0.203733\\
   FCI&-7.221558 & -7.400238&0.178680& -24.252889&-24.435329& 0.182440 \\
   UCCSD &-7.221558& -7.400238&0.178680& -24.252588&-24.422638& 0.170050\\
   \hline \hline 
   \multicolumn{7}{c}{\textbf{3-21G}}\\ \hline
   HF &-7.187095&-7.381511&0.194416&-24.096376&-24.389634&0.293258\\ 
   FCI&-7.187284&-7.381845&0.194561&-24.153345&-24.431849&0.278504\\
   UCCSD &-7.187283&-7.381802&0.194519&-24.145243&-24.418580&0.273327\\
   \hline \hline 
   \multicolumn{7}{c}{\textbf{6-31G}}\\ \hline
   HF & -7.235480& -7.431235&0.195755& -24.234042& -24.519348&0.285306\\ 
   FCI&-7.235643 & -7.431554&0.195911& -24.293125 & -24.562892&0.269767 \\
   UCCSD &-7.235643& -7.431529&0.195886& -24.280634& -24.549776&0.269142
  \end{tabular}
  \end{ruledtabular}
  \label{table1}
  \end{table*} 

\begin{figure*}[t]
    \centering
    \setlength{\tabcolsep}{1mm}
        \begin{tabular}{cc}
            \includegraphics[height=60mm,width=80mm]{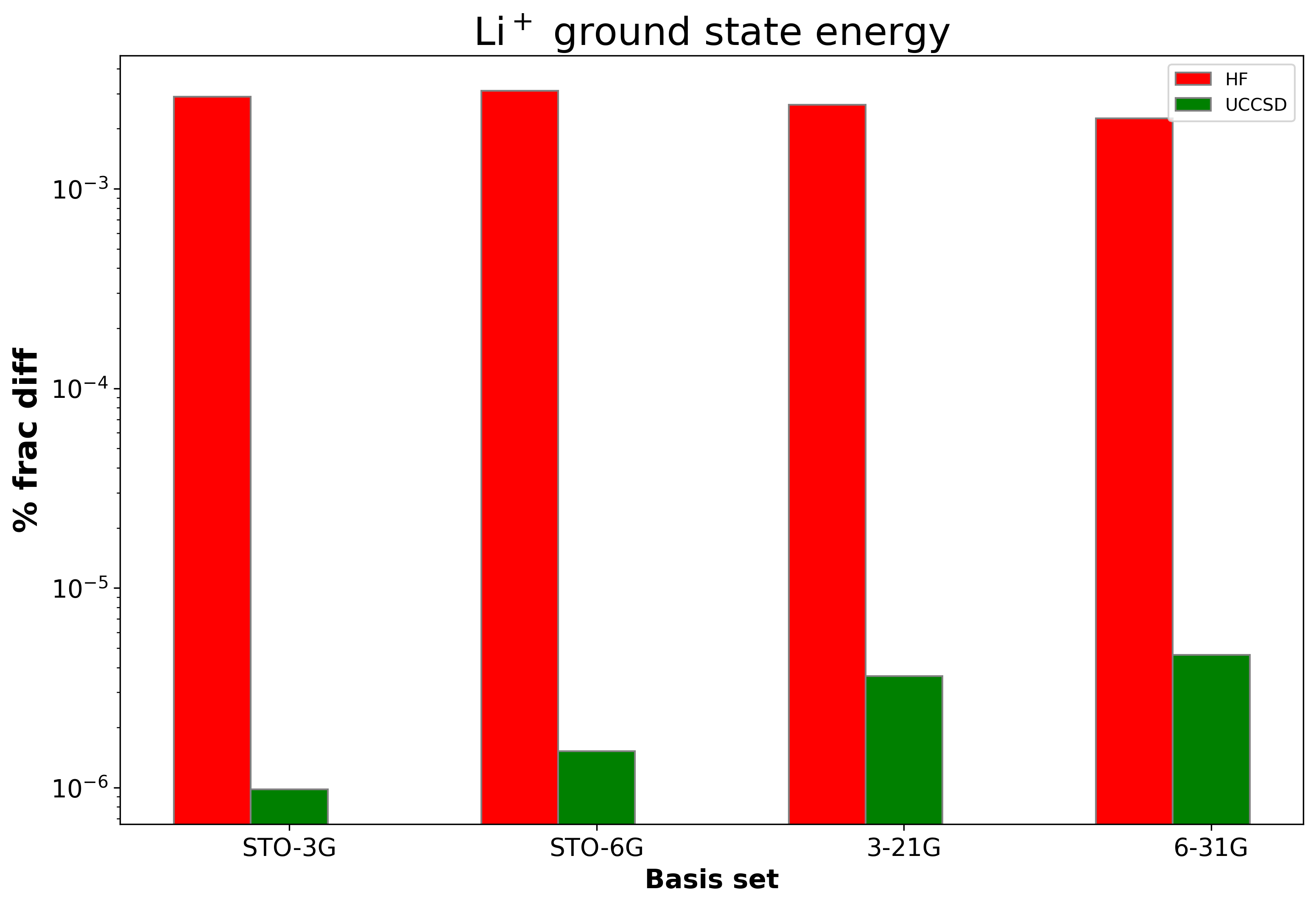} & \includegraphics[height=60mm,width=80mm]{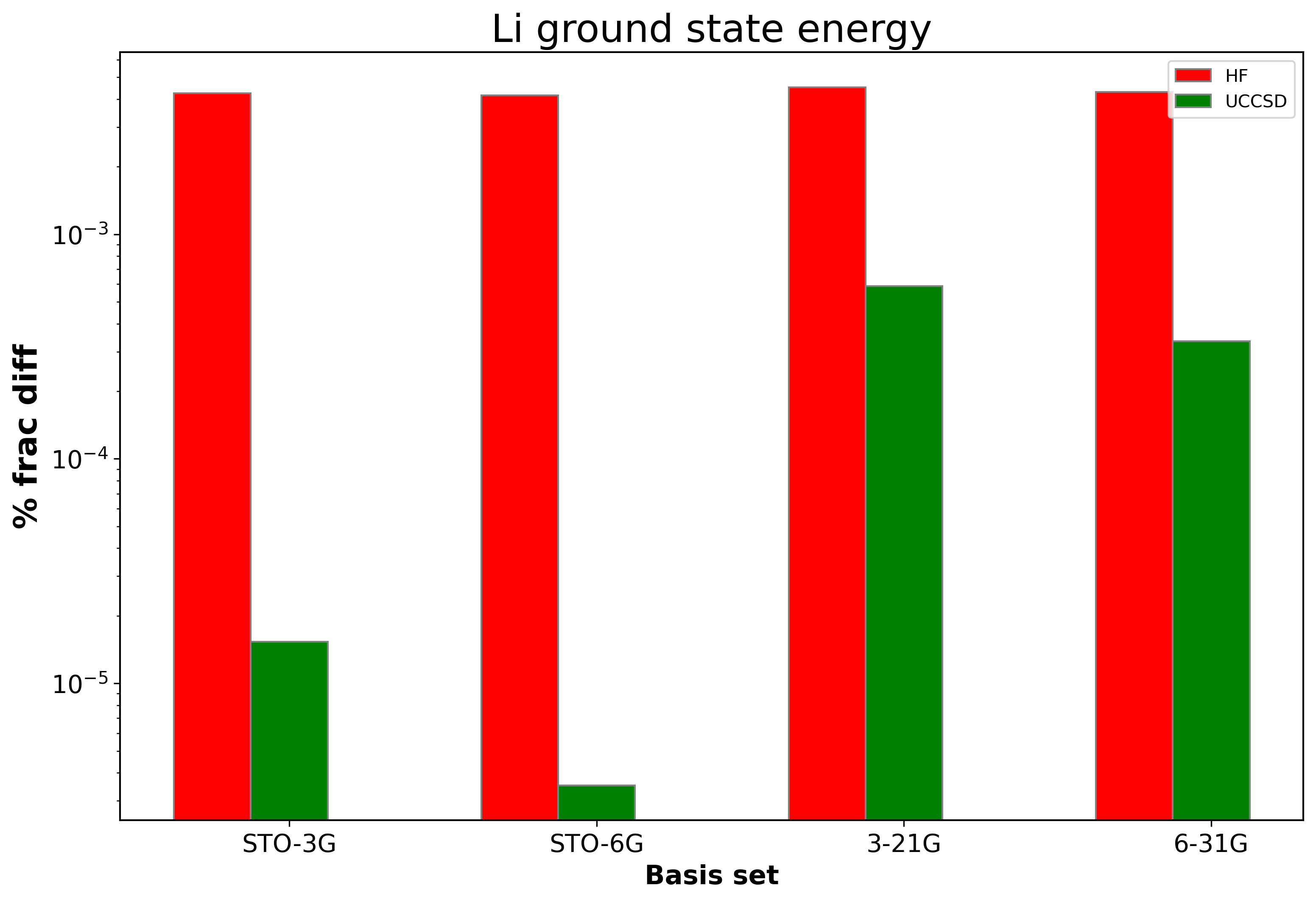} \\
            \textbf{(a)}& \textbf{(b)}\\
            \includegraphics[height=60mm,width=80mm]{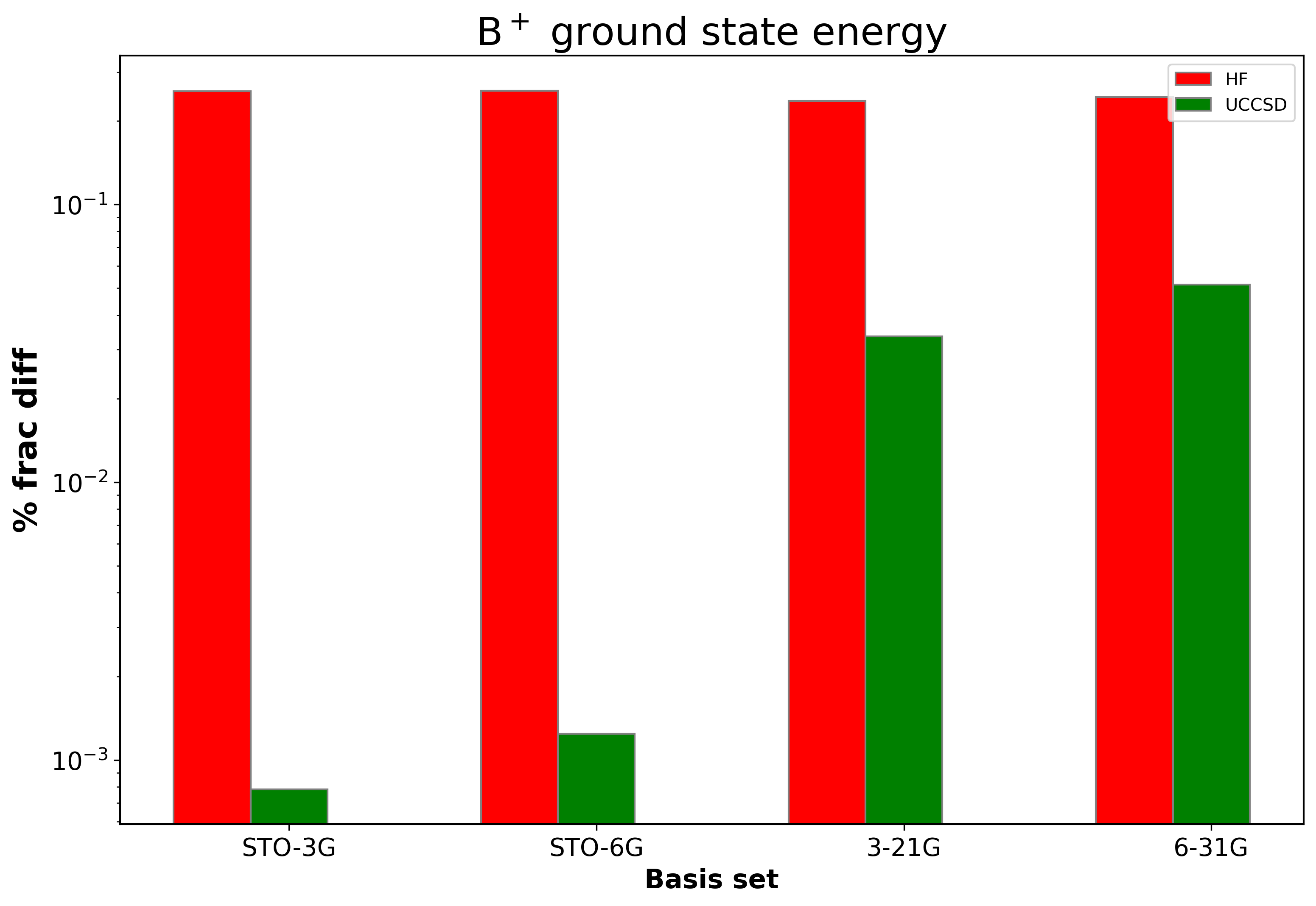} & \includegraphics[height=60mm,width=80mm]{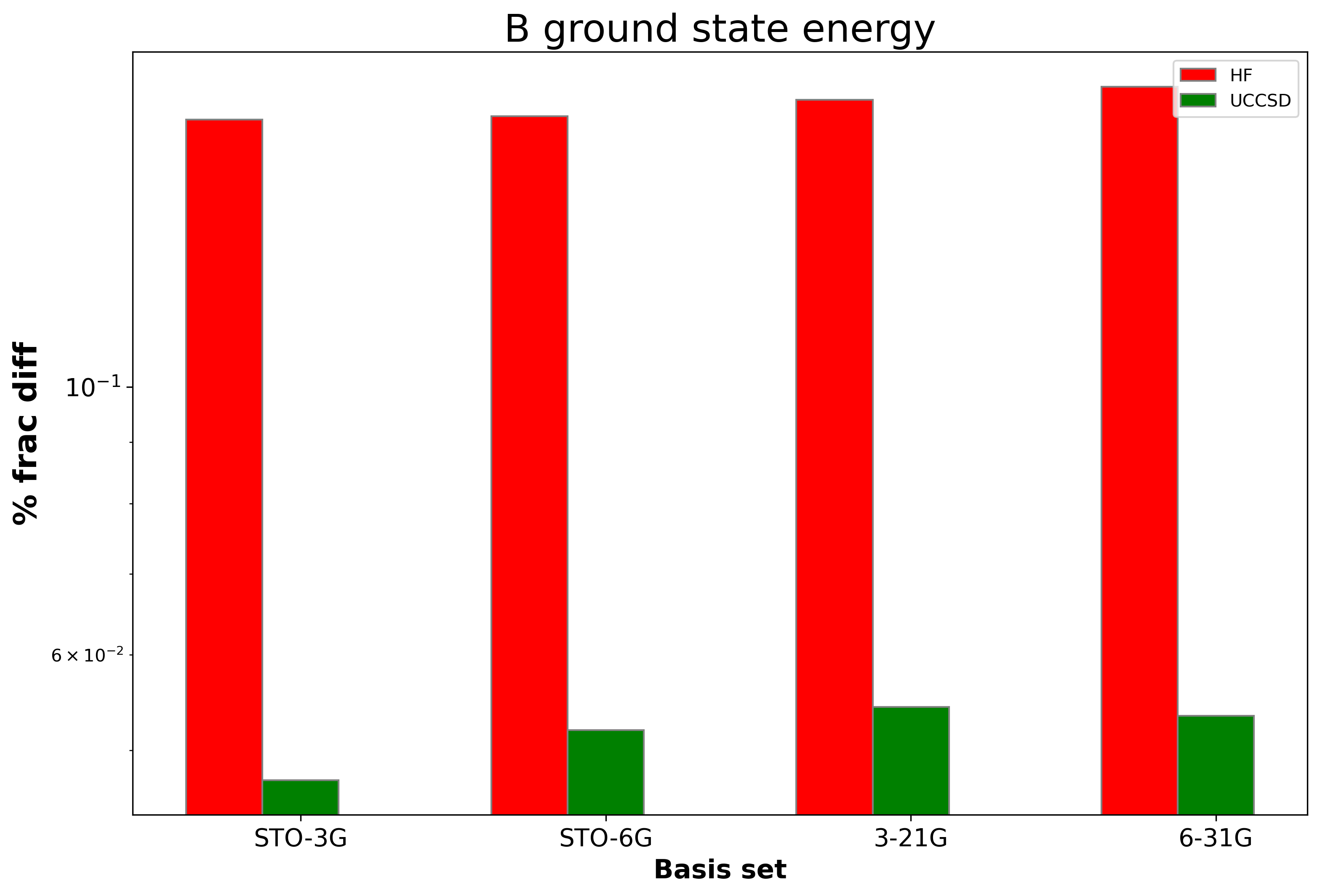} \\
            \textbf{(c)}& \textbf{(d)}\\
        \end{tabular}
\caption{Bar chart showing the percentage fraction difference with respect to FCI (denoted as relative error) for the ground state energies of the neutral and the first ionized counterparts of the two chosen species, $Li$ and $B$, across four basis sets. The red bars correspond to the relative error obtained using the Hartree-Fock method. Green bars correspond to those from the VQE algorithm that uses the UCCSD ansatz. }
\label{fig:E}
\end{figure*}

In this work, we choose the unitary coupled cluster singles and doubles (UCCSD) ansatz for $U(\theta)$. The UCCSD method is the unitary variant of the traditional coupled cluster approach~\cite{ucc}, thus making the latter suitable for quantum computing applications. We reiterate that the coupled cluster theory is the gold standard of electronic structure calculations, due to the highly accurate results that it usually  yields, when compared to other many-body methods at the same level of truncation. Moreover, the theory also possesses desirable theoretical features such as size extensivity~\cite{Helgaker}. Therefore, $\ket{\Psi(\theta)}$ in the UCCSD theory is written as 

\begin{eqnarray}
\ket{\Psi(\theta)} &=& e^{T - T^\dagger}\ket{\Phi_0} \\
&=& e^{\tau} \ket{\Phi_0} ; \\
\tau &=& \tau_1 + \tau_2,\ \mathrm{and} \\
\tau_1 &=& \sum_{ia} \theta_{ia} [a^{\dagger}_a a_i - a^{\dagger}_i a_a], \\
\tau_2 &=& \frac{1}{4}\sum_{ijab} \theta_{ijab} [ a^{\dagger}_a a^{\dagger}_b a_i a_j - a^{\dagger}_j a^{\dagger}_i a_b a_a]. 
\end{eqnarray}

Here, $\ket{\Phi_0}$ is the Hartree-Fock (HF) state (Slater determinant), from which particle-hole excitations arise. We adopt the notation that $\{i, j, \cdots\}$ denote holes (occupied spin-orbitals) while $\{a, b,\cdots\}$ denote particles (unoccupied ones). Therefore, $a_a^\dagger$, for example, means that a particle, $a$, is created. In the singles and doubles approximation, all possible one particle-one hole and two particle-two hole excitations (and de-excitations) are allowed, and these are denoted by $T_1$ ($T_1^\dagger$) and $T_2$ ($T_2^\dagger$), respectively. The amplitudes accompanying the second quantized operators in each term play the role of the parameters, $\theta$, in the VQE algorithm. \\

We now turn our attention to the Hamiltonian, $H$, which is expressed in the second quantized form as 

\begin{eqnarray}
    {H}&=&\sum_{pq}h_{pq}{a}^\dagger_p{a}_q+\frac{1}{2}\sum_{pq,rs}h_{pqrs}{a}^\dagger_p{a}^\dagger_q{a}_s{a}_r \\
    &=& \sum_{\alpha} h_\alpha P_\alpha. 
\end{eqnarray}

Here, $\{p, q, \cdots\}$ denote general spin-orbitals and could be either occupied or unoccupied. $h_{pq}$ and $h_{pqrs}$ are the one- and two- electron integrals, respectively. They are obtained from computations on a classical computer, and are fed as inputs to the VQE algorithm. The last line of the expression given above compactly denotes $H$, with $h_{\alpha}$ denoting the integrals and $P_\alpha$ the second quantized operators. \\

In order to prepare a quantum circuit for $U(\theta) \ket{\Phi_0}$, one needs to transform the operators involved in Eq.~(\ref{Eq2}) to their qubit form. This is done by an appropriate mapping procedure that takes second quantized operators to their qubit counterparts. The transformed operators are a string of tensor product of Pauli operators/gates~\cite{Seeley}. In this work, we use the Jordan-Wigner mapping scheme. Note that the UCCSD ansatz contains an an exponential of sum of operators, and hence one needs to use Trotterization to recast it as a product of exponentials~\cite{trotter}, which enables one to recast the expression as a quantum circuit with a sequence of exponentials expressed in their gate forms. Mathematically, Trotterization can be expressed for sum of operators $A$ and $B$ on an exponential as $e^{A+B} = \lim\limits_{n \to \infty} [e^{A/n}e^{B/n}]^n$, where $n$ is called the Trotter number. We choose a Trotter number of one in this work. We also need to choose appropriately the initial guess parameters, $\theta$. In this work, we begin with zero initial guess for the VQE parameters. Once the circuit for $U(\theta) \ket{\Phi_0}$ is prepared with the initial guess parameters, the energy is calculated (we use the statevector backend throughout in this work), and the outputs are passed to an optimizer, where $E(\theta)$ is minimized as a function of the parameters using a suitable classical algorithm. In this work, we use the COBYLA optimizer~\cite{cobyla,h4, cb1, cb2}, with a maximum of 1000 iterations. A classical optimizer, therefore, processes $E(\theta)$ and provides updated values for the parameters in each iteration, which in turn are used for computing the energy, and so on. The procedure ends when the minimum of $E(\theta)$ is reached within a given threshold. \\ 

We carried out VQE calculations for the closed-shell atomic systems using the Qiskit 0.19 code~\cite{qiskit}. For accommodating one valence electron calculations, we made minor modifications to the program. The one- and two- electron integrals were taken from the PySCF package~\cite{pyscf}. We used the contracted versions of STO-3G, STO-6G~\cite{Boys,sto}, 3-21G, and the 6-31G basis sets~\cite{Pople} for our computations. The first two are specified by $[2s,1p]$ (and hence a 10 qubit computation), while the next two are described by $[3s,2p]$ (and therefore a 18 qubit computation) functions. Within each of the single-particle bases, we compared our results with their full configuration interaction (FCI) values. The FCI approach uses the fact that an arbitrary N-electron wave function can be written as a linear combination of all possible Slater determinants constructed from a single particle basis, that is, as a linear combination of $N$-electron basis functions. Therefore, the FCI result for a system serves as a benchmark to compare the results that we obtain from the VQE  calculations. We carry out the FCI computations using the Pyscf code. \\

\section{Results }\label{results}

Fig.~\ref{fig:E} presents the percentage fraction difference between our results and the FCI values, for the ground state energies of the neutral and the singly ionized species that we have considered, while the energies themselves (and the IEs) are presented in Table~\ref{table1}. The plot also gives the percentage fraction difference between the HF results and the FCI ones, thus giving a visual representation of the degree to which correlation effects are captured in the VQE approach. A red bar therefore denotes the amount of electron correlation itself, whereas the difference between a red and its corresponding green bar indicates the amount of electron correlation that VQE captures with respect to FCI, within that basis. We immediately see from the figure that VQE captures electron correlation effects precisely, noting that the Y-axes are in the log scale. The Y-axes also span different orders in each of the sub-plots, due to the difference in basis. We see that although the energies themselves get closer to the actual (experimental) value for each system from the STO series to the split valence series, the percentage fraction error for the VQE calculations increase from STO series to the split valence series. Qualitatively, this is due to the fact that as the number of qubits (spin-orbitals) increase, there are more possible excitations that we miss within a singles and doubles framework. We observe from the table that for $Li$ and $Li^+$, the difference in energies, with respect to FCI, is always well below $\sim$ 1 milliHartree (mHa) level. This leads to the IEs too agreeing with FCI within $\sim$ 1 mHa. Among the two systems, we see that the energy of the closed-shell atom, $Li^+$, agrees slightly better with FCI than its one-unpaired electron neutral counterpart. This difference becomes very clear in the case of boron, where we observe that for the smaller basis sets (STO series), the cation agrees to less than a mHa with respect to FCI. With the split valence sets, both neutral boron and its cation differ by about 10 mHa. In summary, the precision in our results is well below 1 mHa for lithium and its cation, while it is at most $\sim$ 10 mHa for $B$ and $B^+$. \\ 


We now discuss in more detail our IE results, presented in Table~\ref{table1}. The IE is calculated as the difference between the ground state energies of the neutral and the singly ionized species. If the nuclear charge of an atom is large, the outermost electron is expected to be held tighter. The closer the outermost electron is to the nucleus of the atom, the higher the IE of that atom. In fact, we do observe that we obtain about 0.19 Hartrees for $Li$, and around 0.27 for $B$, with the 6-31G basis. The counterbalancing effects, such as the number of shells is not expected to make much of a difference, as $B$ is only marginally larger than $Li$ in terms of atomic number, and hence shielding effects must be somewhat comparable. In both the cases, the valence orbital (both singly occupied) is ionized, and hence both of them must be comparably stable. Lastly, since the systems are light, relativistic effects should be negligible. The observations are also consistent with the general expectation that since $Li$ is to the left of $B$ in the periodic table, the IE of the latter would be larger than that of the former. It is worth noting here that the STO series does not produce the correct trend. In particular, the ground state energies from the STO-6G basis are lower than those from the 3-21G basis, which may lead to the expectation that the former may perform better than the latter, but when we take the differences in energies, 3-21G actually reproduces the correct trend in IE, and not the STO-6G basis. We also add here that when we go beyond precision and look for accuracy, the IE of $Li$ calculated with the 6-31G basis (0.195886 Ha) differs from experiment at the mHa level (0.198142 Ha~\cite{Liexpt}). On the other hand, the calculated (0.269142 Ha) and experimental (0.304947 Ha~\cite{Bexpt}) values of the IE of $B$ differ by about 35 mHa. This can be attributed to the fact that in $B$ and $B^+$, core-core and core-valence interactions are very strong, in comparison to $Li$ and $Li^+$. In particular, for $B$, the core-core correlation between the $2s$ electrons and the core-valence correlation between the $2s$ and $2p$ electrons. In $B^+$, the dominant correlation contribution is expected to stem from  the two $2s$ electrons. In view of these effects being strong, one would need to go well beyond the split-valence bases in boron and its cation to get accurate results. Also, we note that the split valence bases do not have $d$ functions, and given that a configuration with an $s$ electron excited to $d$ is important \cite{Schaefer}, we miss the correlation effects associated with that configuration with these bases. 
Also, recalling that the implementation of the UCC ansatz has been Trotterized, the effects of  Trotterization could be larger for Boron, thus leading to less accurate results. \\

We verify the possible error due to choice of mapping scheme, by choosing the smaller STO-3G and STO-6G bases as representative cases. In neither case, significant difference was observed between the different qubit mappings for a given species, with the utmost difference occurring at $\sim$ 1 mHa. We found the error due to choice of Trotter number negligible, and well within $\sim$ 1 mHa. \\

\section{Conclusion}\label{conclusion}

In conclusion, we have extended the applicability of the VQE algorithm to calculating first ionization energies (IEs), and have carried out calculations on the $Li$ and $B$ atoms with four basis sets of increasing quality. We find that the VQE algorithm can predict the first IEs well within a milliHartree precision for $Li$, and $\sim$ 10s of milliHartree for $B$. The lower precision in the latter can possibly be attributed to the strong correlation effects between the different $2s$ orbitals as well as the $2s$ and $2p$ orbitals. We find that the STO series basis sets predict the wrong trend in the IE as we move from $Li$ to $B$, while the split valence sets predict the correct trend. We find that the effects of Trotter number greater than one and choice of fermionic operator to qubit operator mapping is not substantial. As quantum technology progresses and more qubits are available, we can use this approach to calculate IEs of heavier atoms and molecules. \\

\section*{Acknowledgements}

We are very thankful to Prof. K. Sugisaki for very useful discussions. The calculations were carried out on National Supercomputing Mission's (NSM) computing resource, `PARAM Siddhi-AI', at C-DAC Pune, which is implemented by C-DAC and supported by the Ministry of Electronics and Information Technology (MeitY) and Department of Science and Technology (DST), Government of India. \\

\end{document}